\documentclass[pre,twocolumn,amsmath,amssymb,superscriptaddress,showpacs]{revtex4}

\usepackage{graphicx}
\usepackage{latexsym}
\usepackage{amsmath}
\usepackage{amssymb}
\usepackage{amsfonts}
\usepackage{bm}

\advance\voffset 1.5cm

\newcommand{\la}{\left<}
\newcommand{\ra}{\right>}
\newcommand{\qvec}{\ensuremath{\underline{q}}}
\newcommand{\rvec}{\ensuremath{\underline{r}}}
\newcommand{\rveccm}{\ensuremath{\underline{r}_\text{cm}}}

\newcommand{\ddiff}{\ensuremath{\text{d}}}

\newcommand{\kB}{\mbox{$k_{\rm B}$}}
\newcommand{\kBT}{\mbox{$k_{\rm B}T$}}

\newcommand{\epsEV}{\ensuremath{\epsilon}}

\newcommand{\df}{d_{\rm f}}
\newcommand{\Siter}{S_i}
\newcommand{\Mgen}{M_g}
\newcommand{\Miter}{M_i}

\newcommand{\Niter}{N_i}
\newcommand{\smax}{s_\text{max}}
\newcommand{\numS}{n_S}
\newcommand{\numM}{n_M}

\newcommand{\Nend}{\ensuremath{N_\mathrm{e}}}
\newcommand{\Rend}{\ensuremath{R_\mathrm{e}}}
\newcommand{\Rseg}{\ensuremath{R_\mathrm{s}}}
\newcommand{\Rgyr}{\ensuremath{R_\mathrm{g}}}
\newcommand{\Rgyrsph}{\ensuremath{R_\mathrm{sp}}}

\newcommand{\wroots}{\ensuremath{w_0(s)}}
\newcommand{\sone}{\ensuremath{\langle s \rangle}}
\newcommand{\stwo}{\ensuremath{\langle s^2 \rangle}}
\newcommand{\Wwiener}{\ensuremath{W_\mathrm{1}}}
\newcommand{\rhohat}{\ensuremath{\hat{\rho}}}

\newcommand{\rhostar}{\ensuremath{\rho_{\star}}}
\newcommand{\Sstar}{\ensuremath{S_\text{$\star$}}}
\newcommand{\Gstar}{\ensuremath{G_\text{$\star$}}}
\newcommand{\epsstar}{\ensuremath{\epsilon_\text{$\star$}}}
\newcommand{\Rstar}{\ensuremath{R_\text{$\star$}}}
\newcommand{\vstar}{\ensuremath{v_\text{$\star$}}}
\newcommand{\wstar}{\ensuremath{w_\text{$\star$}}}

\begin{document}

\title{Hyperbranched polymer stars with Gaussian chain statistics revisited}

\author{P.~Poli\'nska}
\affiliation{Institut Charles Sadron, Universit\'e de Strasbourg \& CNRS, 23 rue du Loess, 67034 Strasbourg Cedex, France}
\author{C. Gillig}
\affiliation{FMF, University of Freiburg, Stefan-Meier-Str. 21, D-79104 Freiburg, Germany}
\author{J.P.~Wittmer}
\email{joachim.wittmer@ics-cnrs.unistra.fr}
\affiliation{Institut Charles Sadron, Universit\'e de Strasbourg \& CNRS, 23 rue du Loess, 67034 Strasbourg Cedex, France}
\author{J. Baschnagel}
\affiliation{Institut Charles Sadron, Universit\'e de Strasbourg \& CNRS, 23 rue du Loess, 67034 Strasbourg Cedex, France}

\begin{abstract}
Conformational properties of regular dendrimers and more general hyperbranched polymer stars with 
Gaussian statistics for the spacer chains between branching points are revisited numerically. 
We investigate the scaling for asymptotically long chains especially for fractal dimensions
$\df = 3$ (marginally compact) and $\df = 2.5$ (diffusion limited aggregation).
Power-law stars obtained by imposing the number of additional arms per generation
are compared to truly self-similar stars.
We discuss effects of weak excluded volume interactions and
sketch the regime where the Gaussian approximation should hold 
in dense solutions and melts for sufficiently large spacer chains.
\end{abstract}

\date{\today}
\maketitle

\section{Introduction}
\label{sec_intro}

\begin{figure}[t]
\centerline{\resizebox{.95\columnwidth}{!}{\includegraphics*{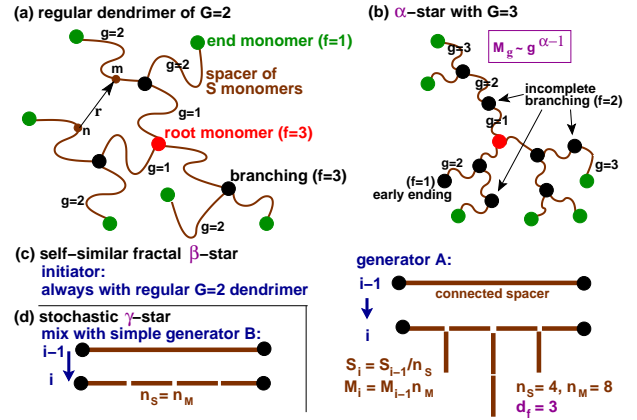}}}
\caption{Sketch of different topologies of branched polymer stars considered:
{\bf (a)}
Regular dendrimer of generation number $G=2$ with $M=9$ arms.
{\bf (b)}
Hyperbranched so-called ``$\alpha$-star" with imposed spacer chain number $\Mgen  \sim g^{\alpha-1}$ 
for $g \le G$ constructed iteratively ($g \to g+1$) by restricting randomly 
the branching of the arms. Some branches may thus end at $g < G$.
{\bf (c)}
Self-similar fractal ``$\beta$-stars" are generated starting with
regular $G=2$ dendrimers and replacing iteratively the $M_{i-1}$ spacers of
length $S_{i-1}$ by $M_i = M_{i-1} \numM$ spacers of length $S_i = S_{i-1}/\numS$.
The generator shown corresponds to self-similar stars of constant density ($\df=3$).
{\bf (d)}
Multifractal ``$\gamma$-stars" are obtained by applying randomly more than one generator.
Mixing with equal weight the generator B ($\numS=\numM=4$) with the compact star
generator A ($\numS=4$, $\numM=8$) leads to a star with $\df=2.5$.
\label{fig_sketch_dend}
}
\end{figure}

\paragraph*{Hyperbranched stars with Gaussian chain statistics.}
Regular exponentially growing starburst dendrimers, as sketched in fig.~\ref{fig_sketch_dend},
and more general starlike hyperbranched chains \cite{foot_nomenetomen} with Gaussian chain statistics
have been considered theoretically early in the literature 
\cite{Stock49,Wiener47,Burchard82,Hammouda92,Biswas94,Obukhov94,foot_df4} 
and have continued to attract attention up to the recent past 
\cite{Blumen03,Blumen09,Blumen09b,Blumen10,Blumen11,Blumen12,Biswas10,Biswas12}. 
One reason for this is that hyperbranched stars \cite{Duplantier89,foot_startheory} 
with sufficiently large spacer chains between the branching points (as indicated by the filled circles) 
are expected to be of direct experimental relevance under melt or $\theta$-solvent conditions
\cite{DegennesBook,DoiEdwardsBook,Duplantier86b}.
Assuming a tree-like structure and translational invariance along the contour, 
the root-mean-square distance $\Rseg$ between two 
monomers $n$ and $m$, as shown in panel (a), is thus given by
\begin{equation}
\Rseg^2 \equiv \la (\rvec_m - \rvec_n)^2 \ra = b^2 s^{2\nu} \mbox{ with }
\nu \equiv 1/2
\label{eq_Rs}
\end{equation}
being the inverse fractal dimension of the spacer chains,
$s$ the curvilinear distance along the tree between both monomers and
$b$ the statistical segment size of the spacer chains \cite{DoiEdwardsBook}.
As a consequence, the typical distance $\Rend$ between the root monomer and the end monomers 
of the most outer generation $g=G$ of spacer chains,
as one possible observable measuring the star size \cite{foot_Rmeasure},
scales as $\Rend^2 = b^2 S G$ with $S$ being the length of the spacer chains (assumed to be monodisperse).
Other moments are obtained from the normalized distribution $P(r,s)$ of the distance $r = |\rvec_m-\rvec_n|$ 
which, irrespective of the specific topology of the branched structure, is given by 
\begin{equation}
P(r,s) = \left(\frac{d}{2\pi \Rseg^2}\right)^{d/2} \exp\left(-\frac{d}{2} \left(\frac{r}{\Rseg} \right)^2 \right)
\label{eq_Grsgauss}
\end{equation}
with $d=3$ being the spatial dimension \cite{foot_dimension}.
Due to their theoretical simplicity such Gaussian chain stars 
(including systems with {\em short-range} interactions along the topological network)
allow to investigate several non-trivial conceptual and technical issues, both for 
static \cite{Blumen09,Blumen11} and 
dynamical 
\cite{Biswas94,Mendoza06,Wu12,Blumen09b,Blumen10,Blumen12} properties, 
related to the in general intricate monomer connectivity imposed by the specific chemical reaction history. 

\paragraph*{Aim of current study.}
We assume here that {\em (i)} the chemical reaction is irreversible (quenched),
{\em (ii)} all spacer chains are monodisperse of length $S$ and
{\em (iii)} flexible down to the monomer scale and 
{\em (iv)} that the branching at the spacer ends is at most three-fold ($f=3$) 
as in the examples given in fig.~\ref{fig_sketch_dend}.
Our aim is 
to revisit various experimentally relevant conformational properties in the limit 
where the total monomer mass $N$ and the total number $M = (N-1)/S$ of spacer chains become 
sufficiently large to characterize the asymptotic {\em universal} behavior and 
to sketch for different star architectures the regimes where the Gaussian spacer chain
assumption becomes a reasonable approximation.
We focus on the large-$S$ limit since this allows under $\theta$-solvent \cite{foot_theta} or melt conditions 
to broaden the experimentally meaningful range of the generation number $G$ of spacer chains.

\paragraph*{Fractal dimension.}
One dimensionless property characterizing the star classes considered
below is their fractal dimension $\df$ which may be defined as \cite{MandelbrotBook,Gouyet}
\begin{equation}
\df \equiv \lim_{R\to \infty} \frac{\log(N)}{\log(R/b)}
\label{eq_dfdef}
\end{equation}
with $N$ being the mass and $R$ the characteristic chain size. 
(Less formally, this definition is often written $N \sim R^{\df}$ \cite{Gouyet}.)
For the regular dendrimers shown in panel (a) the number of spacers $M$ and, hence,
the total mass $N$ increase exponentially with the generation number $G$,
while the typical chain size $R(G) \sim \sqrt{G}$ only increases as a power law. 
That the fractal dimension thus must diverge, is denoted below by the shorthand ``$\df = \infty$".
In addition we shall consider star classes of {\em finite} fractal dimension $\df$,
focusing especially on not too dense systems which should be (at least conceptionally)
of experimental relevance. Specifically, we consider 
{\em (i)} 
marginally compact chains \cite{MSZ11} of fractal dimension $\df = d = 3$ and 
{\em (ii)} 
stars of fractal dimension $\df = 2.5$ which might be thought of as being assembled by 
diffusion limited aggregation (DLA) \cite{Witten81,Meakin83,Meakin86,Gouyet,Mendoza06}.

\paragraph*{Power-law stars.}
As sketched in panel (b), such hyperbranched stars of finite fractal dimension may be constructed 
most readily by imposing a number of spacer chains $\Mgen$ per generation $g$ such that 
the power law $\Mgen \sim g^{\alpha-1}$ holds. Hence, $M \sim N \sim G^{\alpha}$. 
The ``growth exponent" $\alpha$ of these so-called ``$\alpha$-stars" is set by the 
fractal dimension 
\begin{equation}
\alpha = \df \nu
\label{eq_alphadf}
\end{equation}
as may be seen using $N \sim R^{\df}$ and $R \approx \Rend \sim (S G)^{\nu}$ \cite{foot_dimension}.
While being a natural generalization of the regular dendrimer case, restricting the branching of star 
arms does, unfortunately, not lead to a {\em self-similar} tree since the iteration $g \to g+1$ 
is not a proper self-similar generator acting on {\em all} spacer chains \cite{MandelbrotBook,Gouyet}.
We therefore also consider truly self-similar (multi)fractal stars, called in the following
$\beta$- and $\gamma$-stars,
generated iteratively as shown in panel (c) and panel (d) of fig.~\ref{fig_sketch_dend} 
by the iterative application of a well-defined generator (or several generators) 
on {\em all} the spacer chains as in the recent theoretical work on Vicsek fractals \cite{Blumen03}.
For the latter architectures one thus expects to observe for the intramolecular
coherent form factor $F(q)$ the power-law scaling \cite{DegennesBook,BenoitBook,MSZ11}
\begin{equation}
F(q) \sim 1/q^{\df} \mbox{ for } \df \le d =3
\label{eq_Fqdf}
\end{equation}
in the intermediate regime of the wavevector $q$. 
Note that eq.~(\ref{eq_Fqdf}) only holds for open or marginally compact 
self-similar structures \cite{MSZ11,BenoitBook}. 
In fact, Gaussian hyperbranched stars with higher fractal dimension, $\df > d$,
approach with increasing generation number and mass the Gaussian limit
\begin{equation}
F(q) \approx N \exp\left(-(q\Rgyr)^2/d\right) \mbox{ for } q \ll 1/bS^{1/2}
\label{eq_Fq_asympt}
\end{equation}
as shall be demonstrated below.

\paragraph*{Outline.}
The paper is organized as follows:
We summarize first in sect.~\ref{sec_algo} the numerical methods
and specify then in sect.~\ref{sec_topo} the different topologies studied.
Some real space properties are presented in sect.~\ref{sec_real}
before we turn to the characterization of the intramolecular form factor $F(q)$
in sect.~\ref{sec_form}. 
While most of this study is dedicated to strictly Gaussian hyperbranched stars,
i.e. all excluded volume effects are switched off, we investigate more briefly 
in sect.\ref{sec_weakEV} by means of Monte Carlo (MC) simulations \cite{AllenTildesleyBook}  
effects of a weak excluded volume interaction penalizing too large densities.
Even an exponentially small excluded volume is seen to change qualitatively the behavior 
of large regular dendrimers. We conclude the paper in sect.~\ref{sec_conc}.
Neglecting deliberately the long-range correlations expected
as for linear chains \cite{WCX11}, 
we sketch the regime where the Gaussian approximation for melts of hyperbranched
stars should remain reasonable for sufficiently large spacers.


\section{Some computational details}
\label{sec_algo}

\paragraph*{Settings and parameter choice.}
We suppose that the monomers are connected by ideal Gaussian springs. 
The spring constant is chosen such that the effective bond length $b$, eq.~(\ref{eq_Rs}), becomes unity. 
Also, both the temperature $T$ and Boltzmann's constant $\kB$ are set to unity.
All Gaussian spacers are of equal length $S$ (which comprises one end monomer
or branching monomer). With $M$ being the total number
of spacer chains, a hyperbranched star thus consists of $N = 1 + S M$ monomers.
If nothing else is said, $S=32$ is assumed.
(This arbitrary choice is motivated by simulations of dendrimer melts presented elsewhere.)
For $S=32$ we sampled up to a generation number $G=17$ for regular dendrimers
and up to $G \approx 2000$ for power-law hyperbranched stars of fractal dimension $\df=3$ and $\df=2.5$.
(Even larger $G$ obtained using smaller $S$ are included below where appropriate.)
Some properties of the largest system computed for each investigated star architecture
are listed in Table~\ref{tab_Gauss}.

\begin{table}[t]
\begin{tabular}{|c||c|c|c|c|c|c|c|c||}
\hline
star type    & $\df$   & $G$  &$N/10^6$&$\Nend/10^3$
                                              &$\frac{\sone}{\smax}$
                                              &r.f.
                                                    &$\Rend$&$\Rgyr$\\ \hline
Dendrimer    & $\infty$&  17  &12.6    &197 &0.87 &0.11&23     &22     \\
$\alpha$-star& 6       &  50  &22.6    &41.6&0.74 &0.19&40     &34     \\
$\alpha$-star& 5       &  80  &10.4    &10.0&0.70 &0.22&51     &42     \\
$\alpha$-star& 4       & 200  &7.2     &2.2 &0.63 &0.29&80     &64     \\
$\alpha$-star& 3       & 2000 &16.2    &0.4 &0.47 &0.45&253    &138    \\
$\alpha$-star& 2.5   & 2000 &2.4     &0.05&0.36 &0.61&253    &108    \\
$\beta$-star & 3       & 2048 &8.4     &1.2 &0.49 &0.51&256    &179    \\
$\beta$-star & 2.5     & 4096 &1.1     &0.03&0.45 &0.56&362    &171    \\
$\gamma$-star& 2.5     & 8192 &11.1    &0.3 &0.47 &0.54&512    &351    \\
\hline
\end{tabular}
\vspace*{0.5cm}
\caption[]{Various properties for different hyperbranched star types of spacer length $S=32$:
fractal dimension $\df$, largest generation number $G$, total mass $N$, 
number of end monomers $\Nend$ in the last generation shell $g=G$,
rescaled Wiener index $\sone/\smax$ with $\smax = 2 G S$ being the largest curvilinear 
distance between pairs of monomers,
relative root mean-square fluctuation $\sqrt{\stwo-\sone^2}/\sone$ (r.f.) of the normalized histogram $w(s)$,
root mean-square end distance $\Rend$ between the root monomer and the end monomers
of the generation shell $g=G$ and radius of gyration $\Rgyr$. 
\label{tab_Gauss}}
\end{table}

\paragraph*{Local and collective MC moves.}
\begin{figure}[t]
\centerline{\resizebox{0.9\columnwidth}{!}{\includegraphics*{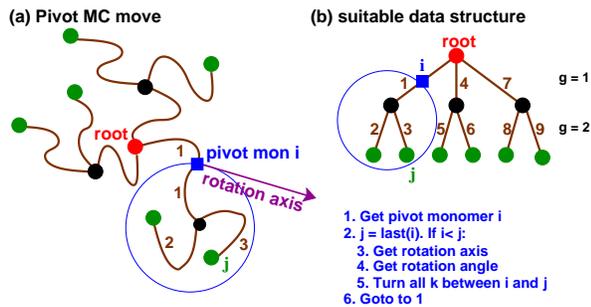}}}
\caption{Sketch of pivot MC move (a) and data structure (b).
A monomer $i$ (filled square) is selected randomly and all attached monomers $k$
closer to the ends (within thin circles) are rigidly turned by an angle $\theta$.
A suitable data structure consists in ordering the
spacer arms (their index indicated by the numbers) and the monomers such that 
all monomers $k$ become neighbors in the monomer lists ($i < k \le j$). 
\label{fig_sketch_algo}
}
\end{figure}
Due to their Gaussian chain statistics many conformational properties can be readily obtained 
using Gaussian propagator techniques \cite{DegennesBook} or equivalent linear algebra relations 
\cite{Blumen09,Blumen12,Biswas10,Biswas11,Biswas12}.
However, some interesting properties, such as the eigenvalues $\lambda_i$ of the inertia tensor,
can be more easily computed by direct simulation which are in any case necessary 
if long-range interactions between the monomers are switched on (see below).
As shown in fig.~\ref{fig_sketch_algo}, we use pivot moves with rigid rotations
of the dangling chain end (as shown by the monomers within the thin circles) 
below a randomly chosen pivot monomer $i$.
The monomers are collectively turned (using a quaternion rotation \cite{AllenTildesleyBook}) 
by a random angle $\theta$ around an also randomly chosen rotation axis through the pivot monomer.
As illustrated in panel (b) of fig.~\ref{fig_sketch_algo}, it is useful to
organize the data structure such that arms and monomers which are turned
together are also grouped together. 
This allows to rotate all monomer $k$ with $i < k \le j$.  
The tabulated monomer $j = last(i)$, the last monomer to be turned, must be an end monomer.
A pivot move does leave unchanged the distances between connected monomers.
(If the connectivity of the monomers is the only interaction, a suggested move is thus always accepted.)
To relax the local bond length distribution simple local MC jumps are added
\cite{AllenTildesleyBook}. The root monomer at the origin never moves.

\paragraph*{Excluded volume interactions.}
Due to excluded volume constraints the volume fraction occupied by a realistic chain can,
obviously, not exceed (much above) unity.
One simple way to penalize too large densities is to introduce 
an excluded volume energy through the lattice Hamiltonian 
\begin{equation}
E = \frac{\epsEV}{2} \sum_{\rvec} n(\rvec) \ (n(\rvec)-1)
\label{eq_Elatt}
\end{equation}
using the monomer occupation number $n(\rvec)$ of a simple cubic lattice.
For all examples presented below we set $\delta x = \delta y = \delta z =1$,
i.e. the grid volume $\delta V = \delta x \ \delta y \ \delta z$ is unity
and $n(\rvec) = \rho(\rvec) \delta V$ measures the instantaneous local density.
The Hamiltonian is similar to the finite excluded volume bond-fluctuation model
for polymer melts on the lattice described in \cite{WCX11,foot_thetasolvent},
however, the particle positions are now {\em off-lattice}
and only the interactions are described by the lattice.
A local monomer or collective pivot move is accepted using the standard Metropolis criterion 
for MC simulations \cite{AllenTildesleyBook}.
Note that the collective pivot moves are best implemented using a second lattice 
for the attempted moves.


\section{Characterization of imposed intrachain connectivity}
\label{sec_topo}

\paragraph*{Introduction.}
We assume that the hyperbranched star topology is not annealed, i.e. not in thermal equilibrium, 
but irreversibly imposed by the chemical reaction.
The first step for the understanding of such quenched structures
is the specification and characterization of the assumed imposed
connectivity, often referred to as ``connectivity matrix" \cite{Blumen09,Biswas12}.
A central property characterizing the monomer connectivity is the normalized histogram
of curvilinear distances 
\begin{equation}
w(s) = \frac{1}{N^2} \sum_{n,m=1}^N \delta(s - s_{nm})
\label{eq_wsdef}
\end{equation}
with $s_{nm}$ being the curvilinear distance between the monomers $n$ and $m$.
Trivially, $w(s=0) = 1/N$ and
$w(s) \approx 2N/N^2 = 2/N$ for $0 < s \ll S$ since the same monomer pair is counted twice.
Note that the histogram $w(s)$, sampled over all pairs of monomers 
of the chain, may differ in general from the similar distribution $\wroots$ of the 
curvilinear distances between the root monomer and other monomers.
We remind also that for a linear polymer chain \cite{WCX11} 
\begin{equation}
w(s) = \frac{2}{\smax} \left(1 - \frac{s}{\smax}\right) 
\mbox{ for } 0 < s \le \smax 
\label{eq_ws_linear}
\end{equation}
with $\smax = N- 1\approx N$.
For most of the star architectures considered the largest curvilinear distance $\smax$
is given by $\smax = 2 S G$.
The histogram $w(s)$ will be used below for the determination of experimentally relevant 
properties such as the radius of gyration $\Rgyr$ and the intramolecular form factor $F(q)$.
The first and second moments of $w(s)$ are given in Table~\ref{tab_Gauss} 
for the different architectures studied. 
We remind that $N \sone$ is sometimes called ``Wiener index" $\Wwiener$ \cite{Wiener47,Biswas12}.

\begin{figure}[t]
\centerline{\resizebox{0.9\columnwidth}{!}{\includegraphics*{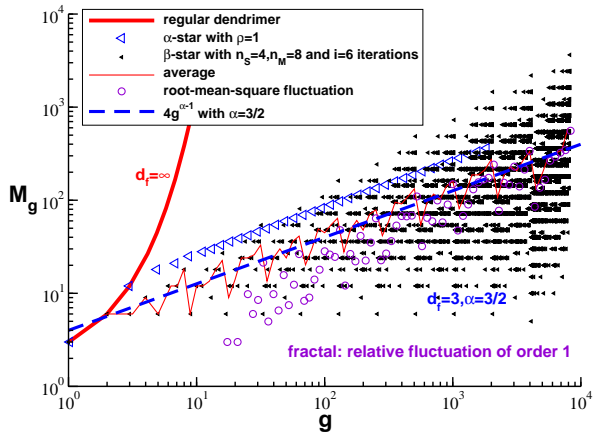}}}
\caption{Number of spacer chains $\Mgen$ for dendrimers (bold solid line) and power-law stars 
of fractal dimension $\df=3$ ($\alpha=3/2$). The open triangles refer to an $\alpha$-star,
the small filled triangles to a $\beta$-star constructed as shown in fig.~\ref{fig_sketch_dend}(c). 
The logarithmically averaged number of arms (thin line) and 
the root-mean-square fluctuations (circles) are of same order.
\label{fig_iRhisto}
}
\end{figure}

\begin{figure}[t]
\centerline{\resizebox{0.9\columnwidth}{!}{\includegraphics*{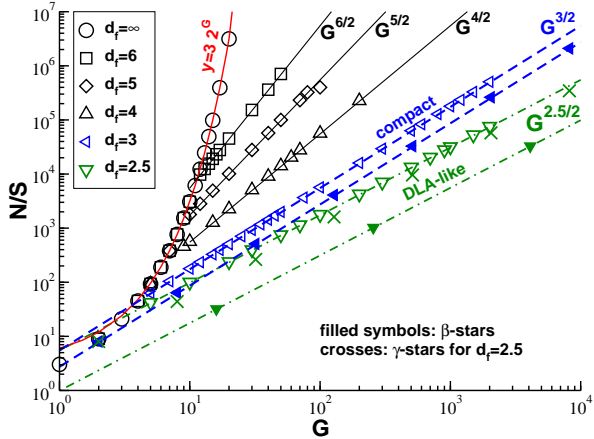}}}
\caption{Number of monomers $N/S \approx M$ {\em vs.} generation number $G$
for different imposed topologies and fractal dimensions $\df$. 
Dendrimers are indicated by $\df=\infty$ (circles), 
$\alpha$-stars by the other open symbols.
The filled triangles corresponds to $\beta$-stars of $\df=3$ ($\numS = 4$, $\numM = 8$) 
and $\df=2.5$ ($\numS = 16$, $\numM = 32$) dimensions, the crosses to $\gamma$-stars of $\df=2.5$.
\label{fig_N}
}
\end{figure}

\begin{figure}[t]
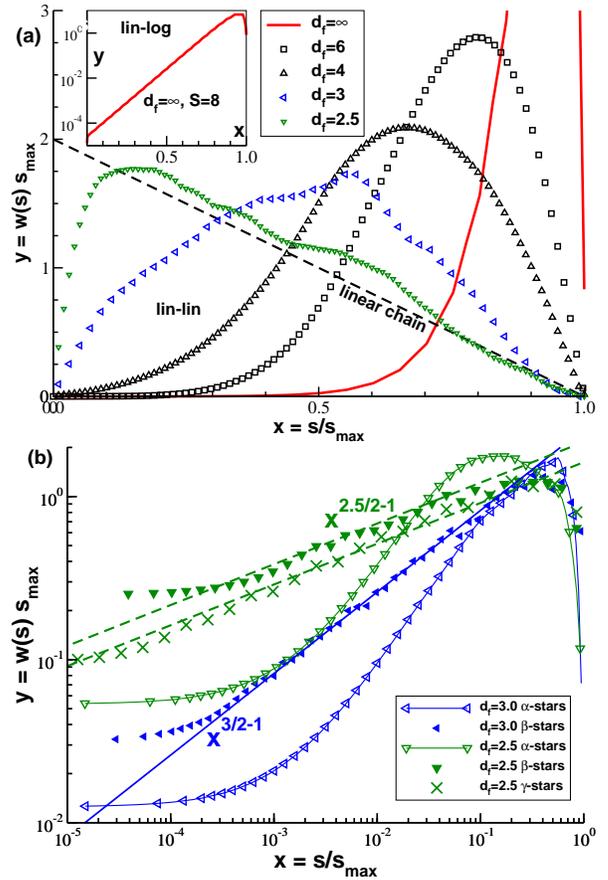

\centerline{\resizebox{0.9\columnwidth}{!}{\includegraphics*{fig5a}}}
\centerline{\resizebox{0.9\columnwidth}{!}{\includegraphics*{fig5b}}}
\caption{Histogram $w(s)$ measuring the number of monomer pairs
at curvilinear distance $s$ along the branched chain:
{\bf (a)} Dendrimers for $G=20$ (bold line) and $\alpha$-stars 
for different fractal dimensions (open symbols). 
The dashed line indicates the histogram for a linear chain of length $N \approx \smax$.
Inset: Half-logarithmic representation for dendrimers.
{\bf (b)} Double logarithmic representation for $\df=2.5$ (upper data)
and $\df=3$ (lower data). As emphasized by the solid and dashed lines
a power law $x^{\alpha-1}$ is observed only for $\beta$- and $\gamma$-stars.
\label{fig_ws}
}
\end{figure}

\paragraph*{Regular dendrimers.}
Let us first summarize several simple properties of the regular dendrimers sketched in fig.~\ref{fig_sketch_dend}(a).
As already mentioned above, the number $\Mgen$ of spacer chains per generation shell $g \le G$
increases exponentially as $\Mgen = 3 \cdot 2^{g-1}$ as shown by the bold line in fig.~\ref{fig_iRhisto}.
Since we assume monodisperse spacer chains of length $S$, this implies
$\wroots \approx 2^{s/S}$ for $S \ll s \le S G$ and that the mass $N$ at generation number $G$ 
must also increase exponentially, as shown in fig.~\ref{fig_N}. 
The histogram $w(s)$ of curvilinear distances $s$ for dendrimers is given in panel (a) of fig.~\ref{fig_ws}
(bold solid lines).
The main panel gives a linear representation of the dimensionless rescaled histogram $w(s) \smax$ as function of $s/\smax$,
the inset on the left-hand side a similar half-logarithmic representation. 
As one expects, the histogram increases exponentially for curvilinear distances $S \ll s \ll \smax$ 
due to the exponential increase of alternative paths of length $s$ starting from an arbitrary monomer.
Using simple combinatorics it can be seen that the histogram must become
\begin{equation}
w(s) \approx \frac{2}{N} \ 2^{(s/S-1)/2} \mbox{ for } 1 \le s \ll \smax.
\label{eq_wd_dend}
\end{equation}
The cutoff observed for large $s \approx \smax$ is due to the finite mass of the star
and the finite length of its branches, just as the finite length of a linear chain gives rise to 
eq.~(\ref{eq_ws_linear}).
As seen from Table~\ref{tab_Gauss}, the reduced first moment $\sone/\smax$ approaches unity 
for dendrimers and the relative fluctuations are the smallest for all architectures considered.

\paragraph*{Hyperbranched $\alpha$-stars.}
As already noted in the Introduction, a simple way to generate stars of a finite fractal dimension
$\df$ is to impose a power law $\Mgen = c g^{\alpha-1}$ for the number of spacers per generation shell
with $c$ being a constant \cite{foot_Mgenconst}.
This is done by randomly attaching $\Mgen$ spacer chains to the end monomers of generation $g-1$
(with the constraint that at most two spacers can be attached per end monomer). An example
for such an $\alpha$-star with $\alpha=3/2$ is given in fig.~\ref{fig_iRhisto} (open triangles).
The corresponding total mass $N \approx S G^{\alpha}$ as a function of $G$ is shown for 
$\alpha=6/2$ \cite{foot_df6},
$\alpha=5/2$, $\alpha=4/2$ \cite{foot_df4}, $\alpha=3/2$ and $\alpha=2.5/2$
by open symbols in fig.~\ref{fig_N}.
The histogram $\wroots$ of curvilinear distances from the root monomer increases as 
$\wroots \sim s^{\alpha-1}$ for $S \ll s \le \smax$ as implied by the $\Mgen$-scaling (not shown).
The curvilinear histograms $w(s)$ over all pairs of monomers are presented in the main panel of fig.~\ref{fig_ws}(a).
The histograms are again {\em non-monotonous} increasing first due to the branching
and decreasing finally due to the finite length of the star arms. 
The latter decay becomes the more marked the weaker the branching, i.e. the smaller $\alpha$,
getting similar for the smallest exponent $\alpha=2.5/2$ studied to the linear chain behavior, 
eq.~(\ref{eq_ws_linear}), indicated by the dashed line. 
As better seen from the double-logarithmic representation in panel (b) of fig.~\ref{fig_ws}, $\alpha$-stars {\em cannot}
be described by a simple power law or exponential behavior for $w(s)$ \cite{foot_alphastar}. 

\paragraph*{Self-similar $\beta$-stars.}
This is different for self-similar fractals created starting from a $G=2$ dendrimer 
of spacer length $S_0$ (as specified below) as initiator and iterating a generator 
as the one shown in fig.~\ref{fig_sketch_dend}(c).
At every iteration step $i$ a spacer of length $S_{i-1}$ is replaced by $\numM$ spacers of length 
$\Siter= S_{i-1}/\numS$.
Hence, $\Siter = S_0/\numS^i$, $\Miter= 9 \numM^i$, $\Niter -1 = \Siter \Miter = 9 S_0 (\numM/\numS)^i$
and $G_i = 2 \numS^i$ for, respectively, the spacer length, the number of spacers, the total mass
and the generation number of the star.
Importantly, the arms added laterally to the original spacer can always be distributed such that 
the root-mean square end-to-end distance of the original spacer (filled circles)
still characterizes the typical size of the replaced spacer. 
Since $S_i G_i = 2 S_0$ for the curvilinear distance between the root monomer and 
the end monomers in the largest generation shell $g = G_i$, the typical chain size $R$,
thus remains {\em by construction} constant as we shall explicitly verify in sect.~\ref{sec_real}.    
Note that the spacer length $S_i$ of the final iteration step is set by 
\begin{equation}
S \stackrel{!}{=} S_i = S_0/\numS^i,
\label{eq_Simposed}
\end{equation} 
which fixes the mass $N_0 \approx S_0 \approx S \numS^i$ of the initiator star. 
Using $N_i \sim R^{\df} \sim N_0^{\nu\df}$ this implies 
\begin{equation}
\numM = \numS^{\beta} \mbox{ with } \beta = \df\nu
\label{eq_alphabeta}
\end{equation}
relating thus both numerical constants $\numS$ and $\numM$.
As shown for $\df=3$ ($\numS=2^2$, $\numM=2^3$) by the small filled triangles in fig.~\ref{fig_iRhisto},
such a self-similar construction leads to a strongly fluctuating number $\Mgen$ of spacers. However,
as shown by the thin solid line the (logarithmically) averaged number of arms 
still increases as $\Mgen \sim g^{\alpha-1}$ with $\alpha = \beta = \df \nu$ in agreement
with eq.~(\ref{eq_alphadf}). Interestingly, the corresponding (also logarithmically averaged)
root-mean square fluctuations (as indicated by open circles) are of the same order,
i.e. the relative fluctuations of spacer number $\Mgen$ per generation shell are of 
order one. The important point is here that all monomers are {\em statistically equivalent}
and that the root monomer does not play any specific role which would break the
self-similarity. (As we have verified, this implies $w(s) \approx \wroots$.)
Averaging over all spacer chains, the total mass $N$ scales, as expected,
again as $N/S \approx G^{\alpha}$ with $\alpha=\df\nu$ as shown in fig.~\ref{fig_N}
by filled triangles for $\df=3$ and $\df=2.5$.
The latter architecture, constructed using $\numS = 2^4$ and $\numM = 2^5$,
is motivated by the fractal dimension $\df \approx 2.5$ which may characterize 
self-similar stars generated by DLA in $d=3$ dimensions \cite{Witten81,Meakin83,Meakin86,Gouyet}.
In our view this is one interesting universal limit of (at least conceptional)
experimental relevance \cite{Mendoza06}. Being self-similar all monomers are equivalent
and since the number of monomers at a curvilinear distance $s$ must increase
on average as $(s/S)^{\alpha-1}$, one expects for $S \ll s \ll \smax$ the power-law scaling 
\begin{equation}
w(s) \approx N \times \frac{1}{N^2} (s/S)^{\alpha-1} \approx \frac{1}{\smax} (s/\smax)^{\alpha-1} 
\label{eq_ws_fractal}
\end{equation}
with $N \approx S (\smax/S)^{\alpha} \approx S G^{\alpha}$.
This is confirmed by the histograms (filled symbols) shown in fig.~\ref{fig_ws}(b).

\paragraph*{Stochastic two-generator multifractals.}
Since the DLA limit is of some importance we have sampled a second system class
of fractal dimension $\df=2.5$ constructed by mixing the generator A
for marginally compact stars shown in panel (c) of fig.~\ref{fig_sketch_dend}
with the second generator B shown in panel (d). 
Being constructed using more than one generator these so-called ``$\gamma$-stars" 
are in fact multifractals \cite{Gouyet,Meakin86}.
(We remember that DLA clusters are also multifractal \cite{Meakin86}.
No multifractal analysis \cite{Gouyet} is required here, however.)
For a given spacer we apply the generator A with a probability $f_A$
and the generator B with a probability $f_B=1-f_A$.  
By choosing different values of $f_A$ any fractal dimension between $\df=2$ and $\df=3$ can be sampled
using both generators.
By reworking the arguments leading to eq.~(\ref{eq_alphabeta}) it can be seen that  
$f_A = f_B =1/2$ corresponds to $\df=2.5$.
While $\beta$-stars are deterministic, the $\gamma$-stars have a {\em stochastic} 
topology due to the random mixing of both generators and an ensemble average over 
several stars is thus taken. 
As may be seen from the crosses in fig.~\ref{fig_N} and fig.~\ref{fig_ws}(b),
the properties of $\beta$- and $\gamma$-stars are, however, rather similar.


\section{Real space characterization}
\label{sec_real}

\begin{figure}[t]
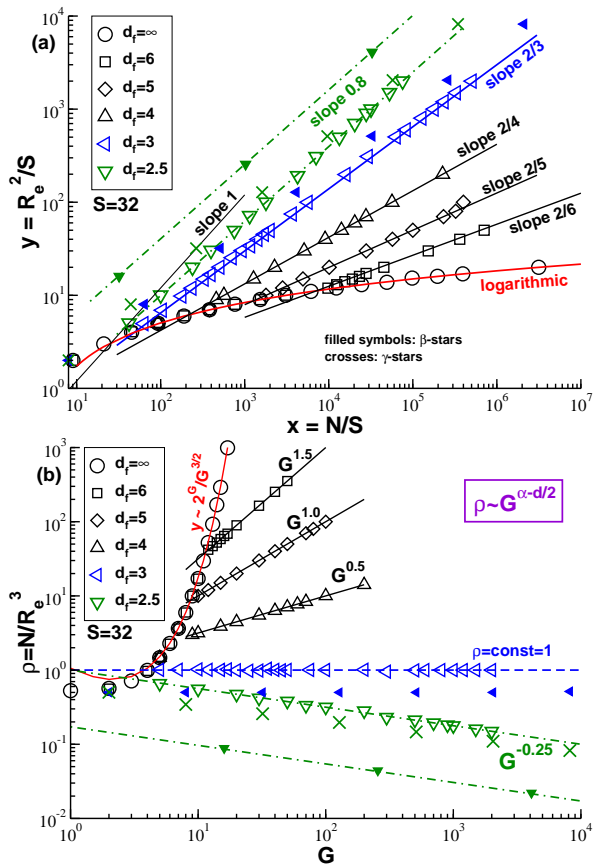

\centerline{\resizebox{0.9\columnwidth}{!}{\includegraphics*{fig6a}}}
\centerline{\resizebox{0.9\columnwidth}{!}{\includegraphics*{fig6b}}}
\caption{Root-mean square end distance $\Rend$ for different imposed topologies:
{\bf (a)}
Double-logarithmic representation of the reduced mean-squared end distance $\Rend^2/S$ {\em vs.} reduced mass $N/S$. 
{\bf (b)}
Density $\rho \equiv N/\Rend^3$ {\em vs.} total generation number $G$ for a spacer length $S=32$. 
\label{fig_Re}
}
\end{figure}

\paragraph*{End distance $\Rend$.}
There are several ways to characterize the typical star size $R$ all being equivalent
from the scaling point of view.
A double-logarithmic representation of the reduced mean-square end distance $\Rend^2/S$
{\em vs.} the reduced mass $N/S$ is presented in panel (a) of fig.~\ref{fig_Re}.
Note that the values of $\Rend$ obtained by direct MC simulations
are within statistical accuracy identical to $\Rend^2 = b^2 S G$.
Both data sets are lumped together.
The regular dendrimer size increases, of course, logarithmically with the mass (circles and bold solid line).
The power-law slopes indicated for finite-$\df$ systems are consistent with the definition $N \sim R^{\df}$.
As one measure of the overall density of a star one may define $\rho \equiv N/\Rend^d$.
(Obviously, a suitable order-one geometrical factor, such as $4\pi/3$, might be included in this definition.)
As can be seen from panel (b) of fig.~\ref{fig_Re}, the density for regular dendrimers 
exceeds already at $G=10$ an unrealistic order of $10$ monomers per volume element.
As indicated by the various power-law slopes, $\rho \sim G^{\alpha-d\nu}$ for  
power-law stars of finite fractal dimension, i.e. the density increases for $\df > d$
and decreases for $\df < d$ as it should \cite{foot_dimension,foot_df4}.

\begin{figure}[t]
\centerline{\resizebox{0.9\columnwidth}{!}{\includegraphics*{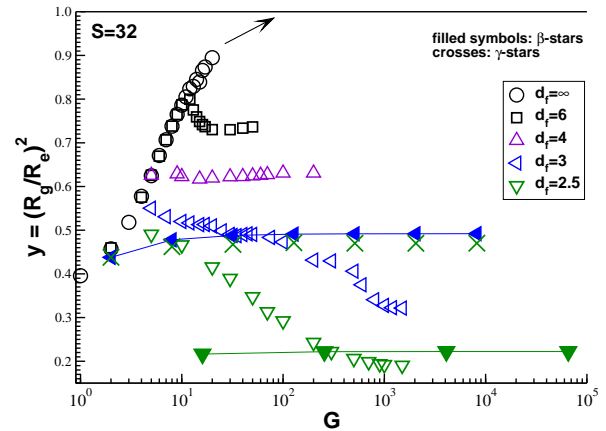}}}
\caption{Reduced radius of gyration $y = (\Rgyr/\Rend)^2$ {\em vs.} generation number $G$.
The ratio $y$ becomes constant only for $\beta$-stars (filled symbols) and $\gamma$-stars (crosses).
\label{fig_Rg}
}
\end{figure}

\paragraph*{Radius of gyration $\Rgyr$.}
The radius of gyration $\Rgyr$ presented in fig.~\ref{fig_Rg} has been determined with 
identical results (lumped again together) either from the MC sampled configuration ensembles 
or by means the formula \cite{RubinsteinBook} 
\begin{equation}
\Rgyr^2 \equiv \frac{1}{2N^2} \sum_{n,m=1}^N \la (\rvec_n-\rvec_m)^2 \ra =
\frac{1}{2} \sum_{s=0}^{\smax} w(s) \Rseg^2
\label{eq_Rg_ws}
\end{equation}
using the histogram of curvilinear distances $w(s)$ discussed above and the
Gaussian chain property $ \Rseg^2 = b^2 s$. Measuring thus the first moment of $w(s)$, 
the radius of gyration is equivalent for Gaussian chains to the Wiener index $\Wwiener$.
The reduced radius of gyration $y = (\Rgyr/\Rend)^2$ is plotted as a function of $G$.
Since the end monomers dominate the mass distribution of regular dendrimers 
for large $G$, $\Rgyr$ becomes similar to $\Rend$. As expected, $y$ approaches unity from below (circles).
Interestingly, the ratio $y$ is constant for the self-similar $\beta$- and $\gamma$-stars,
i.e. $\Rend$ and $\Rgyr$ are similarly rescaled by the iterative application of the generators. 
This confirms the choice of generators discussed in sect.~\ref{sec_topo}.
We note finally that other observables characterizing $R$, 
such as the hydrodynamic radius \cite{DoiEdwardsBook}, have been found to scale 
similarly as the end distance $\Rend$ and the radius of gyration $\Rgyr$. 

\paragraph*{Density profiles.}
\begin{figure}[t]
\centerline{\resizebox{0.9\columnwidth}{!}{\includegraphics*{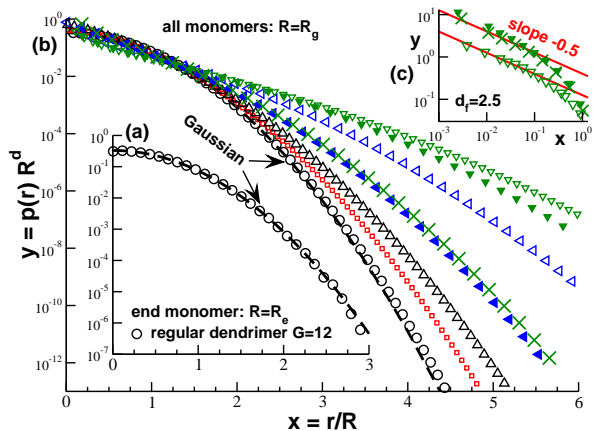}}}
\caption{Density distributions $y = p(r) R^d$ with $r/R$ 
being the reduced distance from the root monomer:
{\bf (a)} 
End monomer distribution with $R=\Rend$ showing the expected Gaussianity (dashed line),
{\bf (b)} 
total monomer distribution rescaled with $R=\Rgyr$
using the same symbols as in fig.~\ref{fig_Rg},
{\bf (c)} 
double-logarithmic representation for three architectures with $\df=2.5$.
The slope indicates the exponent $\df-d=-0.5$. 
\label{fig_pr}
}
\end{figure}
Figure~\ref{fig_pr} presents various normalized density profiles $p(r)$ with $r$ being the 
radial distance from the root monomer. The rescaled distribution $y = p(r) R^d$ is plotted
as a function of the reduced distance $x=r/R$ with $R=\Rend$ in panel (a) and $R=\Rgyr$ in panel (b)
and panel (c). 
The distribution of the end monomers for regular dendrimers ($G=12$, $S=32$) shown in panel (a) 
is a reminder of eq.~(\ref{eq_Grsgauss}), i.e. of the trivial fact that the distances of {\em all} 
pairs of monomers have a Gaussian distribution (dashed line).
The rescaled density $\rho(r) = p(r) N$ of all monomers is shown in panel (b) of fig.~\ref{fig_pr}
(using a half-logarithmic representation) for the largest star of each topology class.
Note that the distribution $p(r)$ has been either obtained for masses up to $N \approx 10^6$
from our MC simulations or for larger systems using 
\begin{equation}
p(r) = \sum_{s=0}^{\smax} \wroots P(r,s)
\label{eq_pr_wroots}
\end{equation}
with $\wroots$ being the already mentioned normalized histogram of monomers
of same curvilinear distance from the root monomer and $P(r,s)$ the size distribution
of a subchain of arc-length $s$ given by eq.~(\ref{eq_Grsgauss}).
Since the density distribution of large regular dendrimers (circles) is dominated by the end monomers, 
$p(r)$ becomes essentially Gaussian (dashed line). We shall come back to this point
at the end of sect.~\ref{sec_form}. The histograms get naturally broader with decreasing $\df$.
Panel (c) on the right-hand side gives a double-logarithmic representation of the
total monomer density distribution for three topologies with $\df=2.5$.
As explained in de Gennes' book \cite{DegennesBook}, the density should decrease as 
$n(r)/r^d \sim 1/r^{d-\df}$ with $n(r) \sim r^{\df}$ being the mass distributed within the volume $r^d$. 
The same power-law exponent is obtained using $\wroots \sim s^{\alpha-1}$ and 
integrating eq.~(\ref{eq_pr_wroots}) for $\df < d$ and $x \ll 1$.
Even the not self-similar $\alpha$-star (open triangles) is seen to follow the predicted slope 
(solid lines). It is sufficient for this property that $\wroots$ has a power-law asymptotics albeit $w(s)$ has not.

\paragraph*{Center of mass fluctuations.}
\begin{figure}[t]
\centerline{\resizebox{0.9\columnwidth}{!}{\includegraphics*{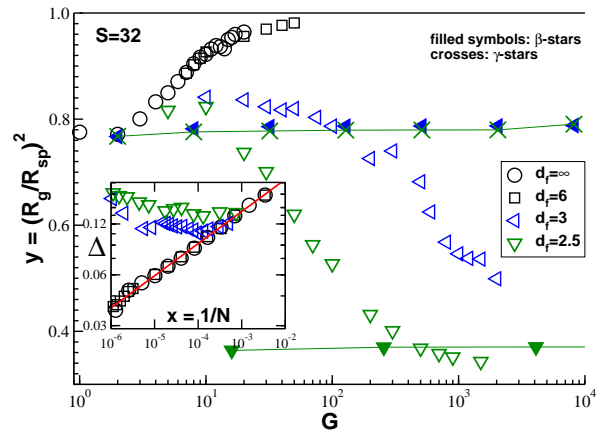}}}
\caption{Aspherical fluctuations: ratio $(\Rgyr/\Rgyrsph)^2$ (main panel) and
rescaled largest eigenvalue $\Delta \equiv \la \lambda_1 \ra/\Rgyr^2-1/3$ of 
the inertia tensor for regular dendrimers and $\alpha$-stars (inset).
\label{fig_sphe}
}
\end{figure}
Albeit spherically averaged density profiles may reasonably characterize {\em some}
aspects of the conformational properties of our hyperbranched polymer stars \cite{Likos11} 
it is important to emphasize that a given instantaneous configuration
may {\em not} be spherically symmetric and depending on the property probed experimentally
or in a computer experiment these aspherical fluctuations become crucial.
This issue is addressed in fig.~\ref{fig_sphe}. The main panel compares the
true radius of gyration $\Rgyr^2 = \frac{1}{N} \sum_n \langle (\rvec_n -\rveccm)^2 \rangle$ 
with a spherical approximation of the mass distribution defined by 
$\Rgyrsph^2 \equiv \frac{1}{N} \sum_n \langle \rvec_n^2 \rangle$ 
assuming the center of mass $\rveccm$ of the star to be set by the
root monomer at the origin for all configurations, i.e. $\rveccm \stackrel{!}{=}0$.
The main panel of fig.~\ref{fig_sphe} presents $(\Rgyr/\Rgyrsph)^2$  as a function of $G$
for different topologies. The ratio is always smaller than unity.
The ratio is seen to approach unity from below for regular dendrimers and $\alpha$-stars with $\df > d$.
While the spherical approximation $\rveccm=0$ becomes thus better with increasing size,
stars with an incredible huge molecular mass are required to reach $\Rgyr \approx \Rgyrsph$.
Interestingly, the ratio {\em decreases} for $\alpha$-stars with $\df = 3 $ and $\df =2.5$ 
(open triangles) while it is essentially constant for the self-similar (multi)fractals. 
For these experimentally most relevant star types the center-of-mass fluctuations remain thus 
relevant for asymptotically large chains.

\paragraph*{Asphericity.}
The asphericity of the stars may be (also) characterized by computing the three eigenvalues 
$\lambda_1 \ge \lambda_2 \ge \lambda_3$ of the inertia tensor of each star and averaging 
over the ensemble. Since $\Rgyr^2 = \langle \lambda_1 \rangle + \langle \lambda_2 \rangle + \langle \lambda_3 \rangle$,
the rescaled eigenvalue $\Delta \equiv \langle \lambda_1 \rangle/\Rgyr^2 -1/3$ should vanish 
for perfectly spherical chains with $\langle \lambda_1 \rangle = \langle \lambda_2 \rangle = \langle \lambda_3 \rangle$.
We have plotted $\Delta$ as a function of the inverse mass for several topologies in the inset of fig.~\ref{fig_sphe}.
As expected from the consideration of $\Rgyrsph$, $\Delta$ is seen to vanish in the large-$N$ limit
for regular dendrimers and $\alpha$-stars with $\df > d$. (As shown by the solid line, $\Delta$
decays only {\em logarithmically} with mass.) 
The opposite behavior is found for smaller fractal dimensions as shown by the open triangles.
Whether for these systems  $\Delta$ becomes constant for $N \to \infty$
(as for linear chains) cannot be confirmed yet from our numerical data.

\section{Form factor}
\label{sec_form}

\paragraph*{Introduction.}
Conformational properties of branched and hyperbranched star polymers can be determined 
experimentally by means of light, small angle X-ray or neutron scattering experiments 
\cite{BenoitBook,Burchard83}. 
Using appropriate labeling techniques this allows to extract the coherent intramolecular form factor 
$F(q)$ defined as
\begin{equation}
N F(q) = \la \rhohat(\qvec) \rhohat(-\qvec) \ra =
\la || \sum_{n=1}^{N} \exp\left(\text{i} \qvec \cdot \rvec_n \right) ||^2 \ra
\label{eq_Fqdef}
\end{equation}
with $\rhohat(\qvec)$ being the Fourier transform of the instantaneous density and
$\qvec$ the wavevector. The average is sampled over the ensemble of thermalized chains.
For sufficiently large $N$ and small $q \equiv ||\qvec||$ the radius of gyration $\Rgyr$, as one 
measure of the star size, becomes the only relevant length scale.
The form factor thus scales as \cite{DegennesBook}
\begin{equation} 
F(q) = N f(Q) \mbox{ with } Q = q\Rgyr
\label{eq_FQscal}
\end{equation}
being the reduced wavevector and $f(Q)$ a universal scaling function with $f(Q) = 1 - Q^2/d$ 
in the ``Guinier regime" for $Q \ll 1$.
The opposite large-$q$ limit probes the density fluctuations within the spacer chains 
and the form factor becomes \cite{DoiEdwardsBook}
\begin{equation}
F(q) = \frac{12}{(b q)^2} \mbox{ for } \frac{1}{bS^{1/2}} \ll q \ll \frac{1}{b}.
\label{eq_Fqlarge}
\end{equation}
For even larger wavevectors correlations on the monomer scale are probed.
In the following we shall focus on the intermediate wavevector range $1/\Rgyr \le q \ll 1/bS^{1/2}$ 
between the Guinier regime and the large-$q$ limit. 

\paragraph*{Dendrimers.}
\begin{figure}[t]
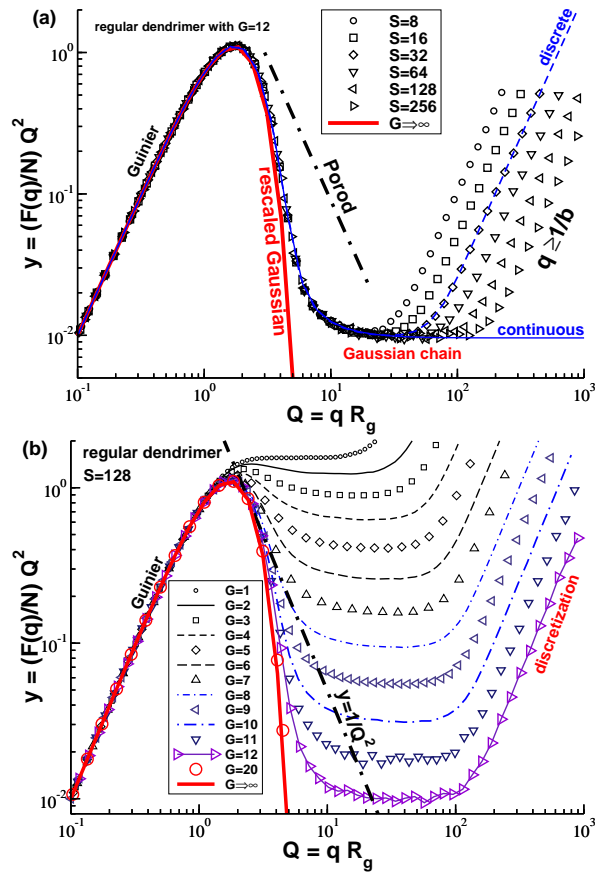

\centerline{\resizebox{0.9\columnwidth}{!}{\includegraphics*{fig10a}}}
\centerline{\resizebox{0.9\columnwidth}{!}{\includegraphics*{fig10b}}}
\caption{Kratky representation of the form factor $y = (F(q)/N) Q^2$ as
a function of the reduced wavevector $Q = q \Rgyr$ for dendrimers:
{\bf (a)} $G=12$ for different spacer length $S$,
{\bf (b)} $S=128$ for different generation number $G$.
The dash-dotted lines indicate the Porod power law \cite{BenoitBook},
the bold solid lines the predicted asymptotic behavior, eq.~(\ref{eq_Fq_asympt}). 
\label{fig_Fq_dend}
}
\end{figure}
Focusing on dendrimers, fig.~\ref{fig_Fq_dend} presents a Kratky representation 
\cite{BenoitBook} of the form factor $y \equiv (F(q)/N) Q^2$ as a function of the 
reduced wavevector $Q = q \Rgyr$. 
Panel (a) shows stars of different spacer length $S$ for a generation number $G=12$,
panel (b) different generation numbers $G$ for a fixed spacer length $S=128$.
The increase of the rescaled data for very large wavevectors $q \gg 1/b$ observed in both panels
is caused by the discrete monomeric units used in our simulations (see below).
The scaling observed for different $S$ in panel (a) for the intermediate wavevector regime,
where the Gaussian spacer chains are probed, is due to the fact that both the mass $N$ and the 
radius of gyration $\Rgyr^2$ are linear in $S$. The corresponding failure of 
eq.~(\ref{eq_FQscal}) in panel (b) shows that there is more than one characteristic length scale.
Note that the strong decay after the Guinier regime above $Q \approx 3$ becomes systematically
sharper with increasing generation number $G$. 
The bold solid lines in both panels indicate the expected asymptotic limit for $G \to \infty$
as discussed at the end of this section. Note that the dendrimer with $G = 20$ 
(large circles) shown in panel (b) is rather close to this limit.
The form factor of this huge chain has not been obtained by MC simulations but by computing numerically 
the equivalent discrete sum
\begin{equation}
F(q) = \sum_{s=0}^{\smax} w(s) P(q,s)
\label{eq_Fq_ws}
\end{equation}
with $w(s)$ being the curvilinear segment histogram discussed above and $P(q,s)$ 
the Fourier transform of the segment size distribution $P(r,s)$.
Since for Gaussian chains $P(q,s) = \exp(-(a q)^2 s)$ with $a \equiv b/\sqrt{2d}$, 
the form factor is readily computed yielding, as one expects, 
the same results as obtained from the explicitly computed configuration ensembles.
This can be seen from the dashed line in panel (a) of fig.~\ref{fig_Fq_dend}
for a spacer length $S=32$.
To compute numerically the form factor using $w(s)$ has the advantage 
that the already mentioned discretization effect at $q \gg 1/b$ can be eliminated.
To do this the discrete sum eq.~(\ref{eq_Fq_ws}) is replaced by a continuous integral 
for $s>0$ and the $s=0$-contribution to the form factor is added.
As shown by the thin solid line in panel (a), this allows to get rid of the 
irrelevant discretization effect. 

\begin{figure}[t]
\centerline{\resizebox{0.9\columnwidth}{!}{\includegraphics*{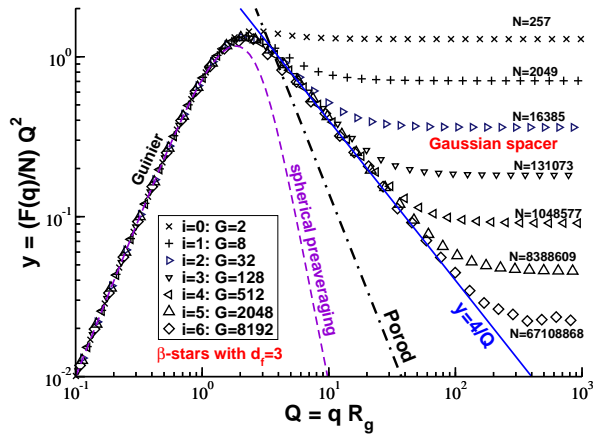}}}
\caption{Kratky representation for $\beta$-stars with $\df=3$.
The reduced form factor approaches with increasing $G$ the power-law slope $-1$ (bold line).
The total monomer mass $N$ is indicated for each iteration $i$.
The dashed line has been obtained according to eq.~(\ref{eq_Fq_preaver}) 
by Fourier transformation of the spherically averaged
density $\rho(r)$ for $i=6$. 
\label{fig_Fq_df3}
}
\end{figure}

\begin{figure}[t]
\centerline{\resizebox{0.9\columnwidth}{!}{\includegraphics*{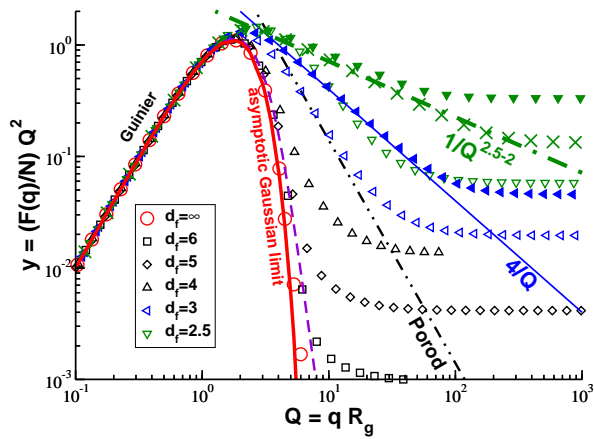}}}
\caption{Rescaled form factor $y(Q) = (F(q)/N) Q^2$ for the largest stars available
obtained using eq.~(\ref{eq_Fq_ws}).
The self-similar $\beta$- and $\gamma$-stars (filled symbols and crosses) decay, as expected,
with a power law $Q^{2-\df}$ in the intermediate wavevector regime as shown by the thin solid line for $\df=3$ ($i=6$) 
and by the dash-dotted line for $\df=2.5$.
The dashed line indicates the preaverage approximation using eq.~(\ref{eq_Fq_preaver})
for $\alpha$-stars of $\df=5$, the bold solid line the expected large-dendrimer limit.
\label{fig_Fq_df}
}
\end{figure}

\paragraph*{Marginally compact stars.}
Figure~\ref{fig_Fq_df3} presents the form factor obtained using the continuous version 
of eq.~(\ref{eq_Fq_ws}) for self-similar fractals of marginal compactness ($\df=3$).
As one expects according to eq.~(\ref{eq_Fqdf}), the data approach with increasing generation 
number the power-law slope $2-\df=-1$ (bold line) expected for the intermediate wavevector regime.
We remind that eq.~(\ref{eq_Fqdf}) can be derived from eq.~(\ref{eq_Fq_ws}) and the
scaling $w(s) \sim s^{\alpha-1}$ for self-similar fractals.
Interestingly, eq.~(\ref{eq_Fqdf}) does {\em not} hold for the (not self-similar)
$\alpha$-stars as may be seen from fig.~\ref{fig_Fq_df}. 
Note also that the large-$q$ plateau of the rescaled form factor in fig.~\ref{fig_Fq_df3} 
only decays as $\Rgyr^2/N \sim 1/N^{1/3}$ extremely slowly with mass. This makes the
numerical confirmation of the power-law slope demanding.
For real experiments this implies that the determination of a
fractal dimension $\df \approx 3$ using the power-law scaling of the form
factor for self-similar stars will also be challenging.
We remind that a similar slow convergence of the intermediate wavevector regime
is well-known for other more-or-less compact polymers
such as polymers confined to ultrathin slits or melts of polymer rings \cite{MSZ11,WMJ13}.

\paragraph*{Comparison of different architectures.}
The rescaled form factors for the largest chains considered for each studied topology are compared
in fig.~\ref{fig_Fq_df}. As expected, all data sets collapse in the Guinier regime below $Q \approx 2$ 
and become again constant for large wavevectors $q \gg 1/bS^{1/2}$. 
(The discretization effect for large $q$ is again avoided using the continuous version of eq.~(\ref{eq_Fq_ws}).)
The decay of the reduced form factor in the intermediate wavevector is seen to become 
systematically stronger with increasing fractal dimension $\df$. 
For the self-similar stars this decay is described by eq.~(\ref{eq_Fqdf}) as emphasized by the solid 
and the dash-dotted power-law slopes for, respectively, $\df=3$ and $\df=2.5$. 
All other architectures decay stronger than a power law. 
Note that it is the shape of this decay which is the most central property to be tested experimentally 
to characterize, at least approximatively, the structure of hyperbranched stars.

\paragraph*{Spherical preaveraging.}
As reminded at the beginning of this section, the intramolecular form factor is the
ensemble average of the squared Fourier transform $\rhohat(\qvec)$ of the fluctuating
instantaneous monomer density.
Following the recent work by Likos {\em et al.} \cite{Likos01},
this begs the question of whether in the limit of large and dense stars,
where density fluctuations should become sufficiently small, 
one may replace $\rhohat(\qvec)$ by the Fourier transform $\rho(\qvec)$ of the 
averaged density profile $\rho(\rvec)$ discussed in sect.~\ref{sec_real}. 
Due to the spherical symmetry of our stars this suggests using 
eq.~(6.54) of ref.~\cite{BenoitBook} the approximation
\begin{equation}
F(q) \approx N \ \left( \int \ddiff\rvec \ p(r) \ \frac{\sin(\qvec \cdot \rvec)}{\qvec \cdot \rvec} \right)^2
\label{eq_Fq_preaver}
\end{equation}
with $p(r) = \rho(r)/N$ being known from eq.~(\ref{eq_pr_wroots}). 
As seen in fig.~\ref{fig_Fq_df3}, eq.~(\ref{eq_Fq_preaver}) 
is not useful for open ($\df < d$) and marginally open ($\df \approx d$) architectures
for which the density fluctuations are yet too large. 
The approximation becomes systematically more successful, however, with increasing fractal dimension
as seen in fig.~\ref{fig_Fq_df} for $\alpha$-stars of fractal dimension $\df=5$.
Note that the striking decay of the rescaled form factor above the Guinier regime
is accurately described by the approximation.
As we have seen in fig.~\ref{fig_pr}, the distibution $p(r)$
becomes systematically more Gaussian with increasing star size and fractal dimension 
since the end monomers of the largest generation shell dominate the total density. 
Since the Fourier transform of a Gaussian is again a Gaussian, 
this implies finally eq.~(\ref{eq_Fq_asympt}) as already stated in the Introduction.
As seen by comparing the solid bold lines in fig.~\ref{fig_Fq_dend} and fig.~\ref{fig_Fq_df}
with the form factors computed using eq.~(\ref{eq_Fq_ws}) for our largest dendrimers (circles),
the asymptotic behavior eq.~(\ref{eq_Fq_asympt}) gives an excellent fit to our numerical data.

\section{Weak excluded volume effects}
\label{sec_weakEV}

\paragraph*{Introduction.}
Up to now we have only considered effects of the imposed monomer connectivity  
assuming all other interactions (persistence length, excluded volume, \ldots) to be switched off. 
Since essentially all properties (apart the eigenvalues $\lambda_i$ of the inertia tensor)
can be obtained analytically or numerically using the Gaussian chain statistics,
the presented MC simulations were less crucial. Direct simulations are, however,
essential for testing the influence of (albeit weak) excluded volume interactions
computed using the lattice occupation number Hamiltonian, eq.~(\ref{eq_Elatt}),
described at the end of sect.~\ref{sec_algo}.

\begin{figure}[t]
\centerline{\resizebox{0.9\columnwidth}{!}{\includegraphics*{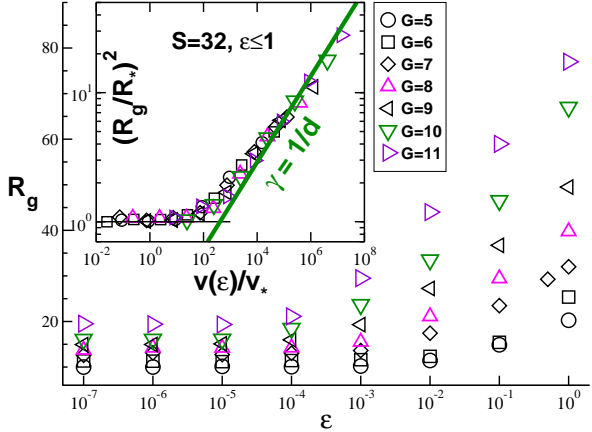}}}
\caption{Radius of gyration $\Rgyr$ for dendrimers 
{\em vs.} excluded volume energy $\epsEV$ for generation number $G$ as indicated. 
Main panel: Unscaled raw data for $S=32$.
Inset: Data collapse of rescaled radius of gyration $(\Rgyr/\Rstar)^2$ as a function of the reduced
excluded volume $v(\epsEV)/\vstar$ with $\Rstar = \Rgyr(\epsEV=0)$ and $\vstar = \Rstar^d/N^2$.
The bold slope corresponds (approximately) to the compact limit $N \sim \Rgyr^d$. 
\label{fig_weakEV_R}
}
\end{figure}

\paragraph*{Scaling of chain sizes.}
Figure~\ref{fig_weakEV_R} presents the excluded volume dependence of the radius of gyration $\Rgyr$ 
for regular dendrimers. (Similar behavior is found for other characterizations of the typical chain size $R$.)
As reveiled in the main panel, the excluded volume effects are the more marked the larger the mass $N(G)$: 
The radius of gyration increases already at $\epsEV=10^{-4}$ for $G=11$ 
while it has barely changed at $\epsEV=0.1$ for $G=5$.
A successful data collapse is seen in the inset of fig.~\ref{fig_weakEV_R}
where the rescaled radius of gyration $(\Rgyr/\Rstar)^2$ is plotted as a function of
the reduced excluded volume $v(\epsEV)/\vstar$ with 
$\Rstar \equiv \Rgyr(\epsEV=0) \approx (S G)^{\nu}$
being the typical size of the Gaussian dendrimer star and
\begin{equation}
v(\epsEV) \equiv \delta V \left(1- \exp(-\beta \epsEV) \right) \approx \beta \epsEV \delta V
\mbox{ for } \beta \epsEV \ll 1
\label{eq_v}
\end{equation}
the excluded volume \cite{DoiEdwardsBook} relevant for our model Hamiltonian 
($\beta$ denoting the inverse temperature). The characteristic excluded volume
$\vstar$ below which the star should remain Gaussian is set by $\vstar \equiv \Rstar^d/N^2$. 
This scaling is a direct consequence of Fixman's general criterion \cite{DoiEdwardsBook}
\begin{equation}
1 \gg v \rho^2 \Rstar^d \approx v N^2/\Rstar^d 
\label{eq_phase_criterion}
\end{equation}
for the Gaussian chain approximation with 
$\rho \approx N/\Rstar^d$ the overall density for Gaussian stars.
That the stars remain Gaussian for $v/\vstar \ll 1$ is emphasized by the horizontal asymptote indicated in the inset. 
The power-law slope $\gamma=1/d$ (bold line) for large reduced excluded volumes
is only an {\em approximative} guide to the eye not taking into account logarithmic corrections.
This can be seen 
{\em (i)} from the usual power-law ansatz \cite{DegennesBook} 
$\Rgyr^2 \approx \Rstar^2 \left( v(\epsEV)/\vstar \right)^{\gamma}$,
{\em (ii)} neglecting the weak logarithmic $N$-dependence of $\Rstar$ (fig.~\ref{fig_Re})
and {\em (iii)} assuming that the dendrimers become essentially marginally compact, $N \sim \Rgyr^d$,
for large $\epsEV$ in agreement with ref.~\cite{Grest96}.
The latter point has explicitly been checked.
For finite-$\df$ stars a similar scaling has been found (not shown). 

\paragraph*{Spacer chain length criterion.}
We note finally that in terms of the generation number $G$ and the spacer length $S$, 
Fixman's criterion may be rewritten remembering that $N \approx S \ 2^G$ for dendrimers
and $N \approx S \ G^{\df \nu}$ for power-law stars \cite{foot_Mgenconst}. 
Hence, the Gaussian approximation must hold for $S \ll \Sstar$ with an
{\em upper} critical spacer length \cite{foot_dimension}
\begin{eqnarray}
\Sstar & \approx & \left( (v/b^d) \  \ 2^{2G}/G^{d\nu} \right)^{-1/(2-d\nu)} \mbox{ and } \nonumber \\
\Sstar & \approx & \left( (v/b^d) \  G^{2\alpha-d\nu} \right)^{-1/(2-d\nu)},
\label{eq_phase_df}
\end{eqnarray}
respectively, for dendrimers ($\df=\infty$) and finite-$\df$ hyperbranched stars.
In both cases $\Sstar \approx (b^3/v)^2$ in $d=3$ dimensions
(while four-dimensional stars are only marginally swollen).

\section{Conclusion}
\label{sec_conc}

\paragraph*{Summary.}
We have revisited by means of direct analytical calculation, using for instance eq.~(\ref{eq_Fq_ws}),
and MC simulations (sect.~\ref{sec_algo})
several conformational properties of regular (exponentially growing) dendrimers 
and power-law hyperbranched stars (fig.~\ref{fig_sketch_dend}) assuming Gaussian chain statistics ($\nu=1/2$).
As emphasized, a central imposed property is the normalized weight $w(s)$ 
of curvilinear distances $s$ between monomer pairs (fig.~\ref{fig_ws}).
Focusing on experimentally measurable observables such as the radius of gyration $\Rgyr$ 
(fig.~\ref{fig_Rg}) and the intramolecular form factor $F(q)$ (figs.~\ref{fig_Fq_dend}-\ref{fig_Fq_df}),
we investigated the scaling for asymptotically long stars with different fractal dimensions $\df$.
Due to their topological simplicity regular dendrimers ($\df=\infty$) have played a central role 
in our presentation (fig.~\ref{fig_Fq_dend})  as in other recent computational studies 
\cite{deGennes83,Muthu90,Grest96,Davies00,Davies01,Lyulin09,Likos01,Likos11}. 
Being (in our view) experimentally and technologically more relevant, 
we have also focused on stochastic architectures with $\df=3$ 
(marginally compact) and $\df = 2.5$ as expected for stars created by DLA \cite{Gouyet}.
We compared ``$\alpha$-stars" constructed by imposing $\Mgen \sim g^{\alpha-1}$ arms per generation
with truly self-similar so-called ``$\beta$-stars" and ``$\gamma$-stars" for which $\Mgen$ 
becomes a strongly fluctuating quantity (fig.~\ref{fig_iRhisto}). As shown in fig.~\ref{fig_Fq_df}, 
only the latter two topologies show the power-law decay, eq.~(\ref{eq_Fqdf}), of the form factor in the 
intermediate wavevector regime expected for open self-similar systems \cite{DegennesBook,BenoitBook,MSZ11}.
While large compact ($\df > d$) stars may roughly be seen as dense colloidal spheres
in agreement with Likos {\em et al.} \cite{Likos01}, the instantaneous aspherical 
fluctuations cannot be neglected for experimentally relevant properties
for the smaller fractal dimensions studied (fig.~\ref{fig_sphe},
dashed line in fig.~\ref{fig_Fq_df3}).
We have commented briefly on the effects of gradually switching on an excluded volume potential.
Coupling the (off-lattice) monomers by means of a (lattice) MC scheme (sect.~\ref{sec_algo}), 
we have sketched for different architectures the regime
($\epsEV \ll \epsstar$, $S \ll \Sstar$) where the Gaussian star approximation can be 
assumed to be reasonable (fig.~\ref{fig_weakEV_R}).

\begin{figure}[t]
\centerline{\resizebox{0.9\columnwidth}{!}{\includegraphics*{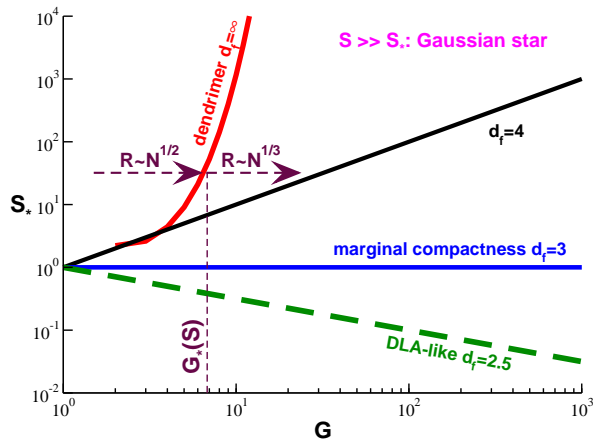}}}
\caption{Sketch of critical spacer length $\Sstar$ for melts.
The Gaussian star assumption holds above the bold lines.
Note that the scaling argument does not allow to fix the scale of the vertical axis.
If the generation number $G$ is increased at constant spacer length $S$, 
as indicated by the dashed arrows, ideal chain behavior is expected for
small $G \ll \Gstar(S)$, while the star becomes colloid-like for larger $G \gg \Gstar(S)$.
The number of chains interacting with a reference star should have
a maximum at $\approx \Gstar(S)$.
}
\label{fig_conc}
\end{figure}

\paragraph*{Conjectures for melts of hyperbranched stars.}
As already pointed out, the Gaussian star assumption should be relevant under melt
conditions assuming a large spacer length $S \gg \Sstar$. That this holds
can be seen by rewriting Fixman's Gaussian chain criterion, eq.~(\ref{eq_phase_criterion}), 
for melts
\begin{equation}
1 \gg \frac{v}{N} \ \rho^2 \Rstar^d \approx v \ N/\Rstar^d
\label{eq_Fixman_melt}
\end{equation}
remembering that the bare excluded volume $v \sim \epsEV$ has to be rescaled by the total 
chain mass $N$ \cite{DegennesBook,WCX11,foot_thetacond}. The hyperbranched stars should thus
remain Gaussian for interaction energies $\epsEV \ll \epsstar \approx \kBT \Rstar^d/(N \delta V)$.
Since $\epsEV$ is not a parameter which can be readily tuned experimentally over several orders of magnitude, 
it is of some importance that eq.~(\ref{eq_Fixman_melt}) sets equivalently a {\em lower} bound $\Sstar \ll S$ 
depending on the generation number $G$.
Following the discussion at the end of sect.~\ref{sec_weakEV},
this implies
\begin{eqnarray}
\Sstar & \approx & \left( 2^G/G^{d\nu} \right)^{1/(d\nu-1)} \mbox{  for } \df=\infty \mbox{ and } \nonumber \\
\Sstar & \approx & \left( G^{(\df-d)\nu} \right)^{1/(d\nu-1)} \mbox{ for finite-$\df$ stars.} 
\label{eq_phasemelt}
\end{eqnarray}
This scaling prediction is sketched in fig.~\ref{fig_conc} for several architectures.
Hyperbranched stars should remain thus Gaussian (albeit with a renormalized effective statistical
segment length \cite{Muthu82,DoiEdwardsBook,WCX11}) as long as $S \gg \Sstar$, if the interaction parameter 
$\beta \epsEV$ is switched on as in the recent study of linear chain polymer melts \cite{WCX11}.
Details may differ somewhat, of course, since the spacer chains may not be rigorously
Gaussian due to long-range correlations related to the overall incompressibility of the melt
\cite{WCX11}. It is thus possible that even self-similar stars of imposed
$\df=2.5$ for the Gaussian reference ($\epsEV=0$) may swell somewhat.
We do conjecture, however, that this ``swelling" for interacting large-$S$ hyperbranched stars
in the melt remains {\em perturbative} as long as $\df < d=3$ \cite{foot_perturbative}.
Considering the dynamical properties of strongly interpenetrating hyperbranched stars
for $S \gg \Sstar$ sampled using standard molecular dynamics \cite{AllenTildesleyBook}, 
it will be of some interest to characterize the mean-square displacement
of the star center of mass or, even better, the associated displacement correlation function
\cite{WCX11}. As for the center of mass motion of linear polymer melts \cite{WCX11,ANS11a},
strong deviations from the Rouse scaling are to be expected \cite{foot_reptation}.

\begin{acknowledgments}
P.P. thanks the IRTG Soft Matter for funding.
We are indebted to A. Blumen and C. Friedrich (both, Freiburg) and
A.~Johner (ICS, Strasbourg) for helpful discussions.
\end{acknowledgments}


\begin{thebibliography}{56}
\bibitem[{foo({\natexlab{a}})}]{foot_nomenetomen}
\bibinfo{note}{For simplicity, we call ``stars" or ``branched stars" all
  tree-like branched architectures, ``regular dendrimers" the deterministic and
  exponentially growing stars shown in panel (a) of fig.~\ref{fig_sketch_dend},
  and ``power-law stars" hyperbranched stars of finite fractal dimension
  $\df$.}
\bibitem[{foo({\natexlab{b}})}]{foot_df4}
\bibinfo{note}{Hyperbranched stars or trees with fractal dimension $\df=4$ have
  been discussed in the context of randomly branched polymers (often called
  ``lattice animals") \cite{Stock49}, dilute rings in a gel of topological
  obstacles \cite{Obukhov94} and as a possible model describing the topological
  interactions of unconcatenated melts of rings \cite{Obukhov94,MSZ11}. All
  these models have in common that on the local scale the branched structure is
  described by Gaussian spacer chains.}
\bibitem[{foo({\natexlab{c}})}]{foot_startheory}
\bibinfo{note}{Stars play a central role in polymer theory in general since the
  partition function of a sufficiently large polymer of any topology can be
  decomposed as a product of a star and a ring partition function as shown by
  Duplantier \cite{Duplantier89}.}
\bibitem[{foo({\natexlab{e}})}]{foot_dimension}
\bibinfo{note}{Although we focus on three-dimensional systems the general
  dimension $d$ is often indicated since this may help to clarify the structure
  of the relations. The reader is invited to replace $d$ by $d=3$. For similar
  reasons we often make explicit the inverse fractal dimension $\nu$ of the
  spacer chains. It should be replaced by its value $\nu=1/2$.}
\bibitem[{foo({\natexlab{f}})}]{foot_theta}
\bibinfo{note}{Due to effective three-body monomer interactions strongly
  branched stars may yet swell in the $\theta$-regime. In the limit of large
  spacer chains this effect should become negligible, however, as for
  unbranched stars for which it has been shown that the renormalized
  interaction vanishes in the limit of infinitely long chains
  \cite{Duplantier86b}.}
\bibitem[{foo({\natexlab{g}})}]{foot_thetasolvent}
\bibinfo{note}{Similarly, one may simulate $\theta$-solvent stars imposing \\
  $E = \epsEV/6 \ \sum_{\rvec} n(\rvec) \ (n(\rvec)-1) \ (n(\rvec)-2)$.}
\bibitem[{foo({\natexlab{h}})}]{foot_Mgenconst}
\bibinfo{note}{The constant $c$ is assumed here to be $S$-independent. Imposing
  instead $\Mgen = \tilde{c} ( S g)^{\alpha-1}$ with $\tilde{c} \sim S^0$ leads
  to a total mass $N \approx (S G)^{\alpha}$ and, hence, to a typical density
  $\rho \approx N/R^d \approx (S G)^{\alpha-d\nu}$. The latter choice may,
  e.g., be useful to compare marginally compact $\alpha$-stars at constant
  density for different spacer lengths.}
\bibitem[{foo({\natexlab{i}})}]{foot_df6}
\bibinfo{note}{A fractal dimension $\df=6$ would arise if a marginally compact
  star is constructed assuming {\em rigid} spacer chains which are then allowed
  to become flexible.}
\bibitem[{foo({\natexlab{k}})}]{foot_thetacond}
\bibinfo{note}{A similar criterion $1 \gg \beta w \rho^3 \Rstar^d$ should hold
  for $\theta$-solvent conditions \cite{foot_thetasolvent} with $w > 0$ being
  the scale of the three-body interactions. Gaussianity thus requires $w \ll
  \wstar \approx \Rstar^{2d}/N^3$. Note that the upper critical dimension for
  $\theta$-solvents being $d=3$, this condition does not depend (up to
  logarithmic corrections) on the spacer chain length $S$.}
\bibitem[{foo({\natexlab{l}})}]{foot_perturbative}
\bibinfo{note}{Basically, each chain builds up a concentration $\rhostar
  \approx N/R^d$ around its center of mass. Monomers belonging to other chains
  are depleted from this region, since the total density $\rho \gg \rhostar$
  can be assumed to be constant. This constraint causes an effective repulsion
  energy $\beta U \approx \rhostar/\rho \sim N^{1-d/\df}$ between chains and
  chain segments \cite{DegennesBook,WCX11}. The ensueing swelling becomes
  perturbative for $1 > \df/d$ \cite{WCX11}.}
\bibitem[{foo({\natexlab{m}})}]{foot_reptation}
\bibinfo{note}{Due to ``viscoelastic hydrodynamic interactions" \cite{ANS11a}
  this should happen even if reptational effects \cite{DoiEdwardsBook} can be
  neglected as is the case for a finite $\epsEV \approx 1$.}
\end{thebibliography}
\end{document}